\documentclass[twocolumn,epsf,graphics,psfig,superscriptaddress]{revtex4}

\usepackage{amsfonts}
\usepackage{graphicx}
\usepackage{amsmath}
\usepackage{amssymb}
\usepackage{latexsym}
\usepackage{psfrag}
\usepackage{dcolumn}
\usepackage{datetime}
\usepackage{multirow}
\usepackage{subfigure}

\begin{document}
\title{Electronic and optical spectra in a diluted magnetic semiconductor multilayer}
\author{Leili Gharaee}
\affiliation{Department of Physics, Payame Noor University, P.O.
Box 19395-3697 Tehran, Iran}
\author{Alireza Saffarzadeh}
\affiliation{Department of Physics, Payame Noor University, P.O.
Box 19395-3697 Tehran, Iran} \affiliation{Department of Physics,
Simon Fraser University, Burnaby, British Columbia, Canada V5A
1S6}
\date{\today}

\begin{abstract}
The effects of random distribution of magnetic impurities with
concentration $x$ in a semiconductor alloy multilayer at a
paramagnetic temperature are investigated by means of coherent
potential approximation and tight-binding model. The change in the
electronic states and the optical absorption spectrum with $x$ is
calculated for weak and strong exchange interactions between
carrier spins and localized spin moments on magnetic ions. We find
that the density of states and optical absorption are strongly
layer-dependent due to the quantum size effects. The electronic
and optical spectra are broadened due to the spin fluctuations of
magnetic ions and in the case of strong exchange interaction, an
energy gap appears in both spectra. Furthermore, the interior
layers show higher contribution in the optical absorption of the
system. The results can be helpful for magneto-optical devices at
a paramagnetic temperature.
\end{abstract}
\maketitle

\section{Introduction}
Semiconductor multilayers are materials with attractive electronic
and optical properties and possible application in nanodevices
\cite{Ridley}. Such unique properties which are caused by their
dependence on the number of atoms in a confined direction, have
led to a rapid growth on research activities in this field in
recent years. For instance, it has been shown that for
semiconducting multilayer armchair graphene nanoribbon with more
than two layers, semiconductor-to-metallic transition is obtained
with increasing electric field, which can be utilized to enhance
the efficiency of graphene based transistor devices \cite{Kumar}.
In addition, for the thicker systems (i,e. more layers), this
transition occurs at a smaller electric field which makes them to
enhance the on/off ratio of a field-effect transistor device
\cite{Kumar}.

Furthermore, magnetic and nonmagnetic impurities in layered
structures have received considerable attention
\cite{Schiller,Moodera1,Moodera2,Pekareka1,Sankowski,Pekareka2,Lashkarev,Chung,Luo}.
Spin-dependent scattering of charge carriers in magnetic
multilayers is the origin of magnetotransport effects and has been
the subject of intensive interest in recent semiconductor
electronics research
\cite{Pekareka1,Sankowski,Pekareka2,Lashkarev,Chung,Luo}. Doping
of magnetic ions in nonmagnetic semiconductors (NMSs) introduces
not only magnetic moment but also free carries in the system. In
diluted magnetic semiconductors (DMSs), the host material is an
alloy semiconductor of the type II-VI or III-V semiconductor with
characteristic of $AB$, in which the atom $A$ is substituted by a
magnetic atom with concentration of $x$ \cite{Furdyna1,Jungwirth}.
For example, Mn ions as magnetic impurities into GaAs and InAs act
as acceptors in the III-V compound semiconductors
\cite{Furdyna1,Ohno}. These semiconducting alloys are a novel
class of ferromagnetic materials and one of the promising
materials for spintronics \cite{Dietl}. The exchange interaction
between a carrier spin and localized spin of magnetic ions, plays
an important role in magneto-optical effects in DMSs
\cite{Ohno,Ando}. Because this interaction strongly affects on
band splitting that is observed in magneto-optical measurements
such as magneto-reflectivity and magneto-absorption spectra. Thus,
the most powerful tool for studying the exchange interaction
between a carrier and localized spins in DMS-based structures is
optical measurement. Furthermore, because of the ternary nature of
DMSs, one can tune the lattice constants and band parameters by
varying the composition of the material. Therefore, most of the
magnetic semiconducting alloys are excellent candidates for
preparation of quantum wells, superlattices, and other
heterostructures that involve band gap engineering
\cite{Furdyna2,Slobodskyy,Kepa,Kirby}.

In previous studies \cite{Saffar2,Shinozuka1,Shinozuka2}, by
applying the single-site coherent potential approximation
(CPA)\cite{Gonis} to semiconductor alloy multilayers, the effects
of chemical (nonmagnetic) disorder on optical absorption spectrum
were investigated. The results showed that, the photon absorption
by the system is layer dependent and strongly changes by variation
of the scattering-strength of the chemical impurities
\cite{Saffar2}. In the present study, we extend our previous
approach  \cite{Saffar2} for the magnetic impurities at a
paramagnetic temperature, and investigate the electronic and
optical properties of DMS multilayers. To study the optical
absorption spectrum in the presence of the magnetic impurities, we
use the Onodera-Toyozawa theory \cite{Onodera}, which is related
to a binary mixed crystal or an alloy consisting of atoms (or
molecules) $A$ and $B$, in which the transition dipole moments of
$A$ and $B$ atoms are equal. We should point out that, the
application of the Onodera-Toyozawa theory together with the CPA
for DMSs was first employed by Takahashi \cite{Takahashi} in which
the optical band edge was studied for bulk materials. The
calculations, however, were performed only for a simple
semicircular model density of states.

In this study, owing to the substitutional disorder of the
magnetic ions in a DMS multilayer, the effect of carrier
scattering due to the presence of both magnetic and nonmagnetic
atoms will be studied within the single-site CPA. Due to the
quantum size effect in the multilayer systems the electronic
states at each layer strongly depend on the location of that layer
relative to the surface layer. Hence, each layer has a different
response to the carrier scattering process in the layer and should
be taken into account in the CPA equations. Since in a
low-dimensional system many important physical properties and
characteristics such as optical properties, are governed strongly
by the electronic structure of the system, then the local density
of states (LDOS) will be analyzed precisely. Therefore, the
dependence of electronic and optical spectra on the carrier
energy, magnetic impurity concentration, and the strength of
exchange interaction between the carrier spin and the spin of
magnetic ions will be studied with more details. This paper is
organized as follows: we start with introducing the model and
formalism in Sec. II. In Sec. III, the results of the numerical
calculations for DMS multilayer are given. The conclusion is
presented in Sec.IV.

\section{The Model}
In this study, we consider a semiconductor multilayer described by
the single-band tight-binding model with the nearest-neighbor
hopping and the on-site delta-function-like potential on a simple
cubic lattice in which one of the dimensions (the $z$ direction)
is confined. The number of layers along the confined direction is
$N_z$ and the label of each layer is denoted by $n$, hence $1\leq
n\leq N_z$. The multilayer is a semiconducting alloy of the form
$A_{1-x}M_xB$, where the sites of $A$ atoms such as Ga can be
occupied by $M$ atoms such as Mn with concentration $x$ and $B$
atoms such as As remain unchanged. Due to the low carrier density
in the system, the interaction between carriers is ignored in the
calculations. Accordingly, the one-electron or a Frenkel exciton
Hamiltonian in this system is given by:
\begin{equation}\label{1}
H=\sum_{{\bf r},n}u_{{\bf r},n}^{A,M}-\sum_{{\bf r},n}\sum_{{\bf
r'},n'}\sum_{\mu}t_{{\bf r}n,{\bf r'}n'}|{\bf
r},n,\mu\rangle\langle {\bf r'},n',\mu|
\end{equation}
where, $|{\bf r},n,\mu\rangle$ is an atomic orbital with spin
$\mu(=\uparrow$ or $\downarrow)$ at site $({\bf r},n)$. Here,
${\bf r}$ denotes the positional vector in the $x-y$ plane of the
layer $n$. The hopping integral $t_{{\bf r}n,{\bf r'}n'}$ is equal
to $t$ for the nearest neighbor sites and zero otherwise. We
assume that the hopping integral only depends on the relative
position of the lattice sites. Therefore, the type of disorder in
our model is considered to be diagonal and accordingly, the
off-diagonal disorder is outside of the scope of the present
study. The random site energy $u^{A,M}_{{\bf r},n}$ is assumed to
be $u^{A}_{{\bf r},n}$ and $u^{M}_{{\bf r},n}$ with probabilities
$1-x$ and $x$ when the lattice site (${\bf r},n$) is occupied by
the $A$ and $M$ atoms, respectively
\cite{Takahashi,Saffar2,Saffar1}.  For $A$ and $M$ sites we have:
\begin{equation}\label{2}
u_{{\bf r},n}^{A}=\varepsilon_A\sum_{\mu}|{\bf
r},n,\mu\rangle\langle{\bf r},n,\mu|
\end{equation}
\begin{equation}\label{3}
u_{{\bf r},n}^{M}=\sum_{\mu,\nu}|{\bf
r},n,\mu\rangle[\varepsilon_M\delta_{\mu\nu}-I\mathbf{S}_{{\bf
r}n}\cdot{\mathbf\sigma}_{\mu\nu}]\langle {\bf r},n,\nu|
\end{equation}
where $I\mathbf{S}_{{\bf r}n}\cdot{\mathbf\sigma}_{\mu\nu}$ is the
$k$-independent exchange interaction in which $|\mathbf{S}_{{\bf
r}n}|=S$ is the local spin operator of the $M$ atom and $\sigma$
is the Pauli matrix for carrier spin. In this study, instead of
taking the quantum fluctuation of the localized spin of the
magnetic ions into account, the classical spin approximation (i.e.
$S\rightarrow\infty$) is used, while $IS$ is a constant parameter
\cite{Takahashi,Saffar1}.

According to the CPA, a disordered alloy can be replaced by an
effective periodic medium in which the potential of all sites is
replaced by an energy-dependent coherent potential, except one
site which is denoted by impurity \cite{Gonis}. Therefore, the
multilayer Hamiltonian $H$ is replaced by the following effective
medium Hamiltonian:
\begin{equation}\label{4}
H_\textrm{eff}=\sum_{{\bf r}n}\sum_{{\bf
r'}n'}\sum_{\mu}[\Sigma_{n}(\omega)\delta_{{\bf r},{\bf
r'}}\delta_{n,n'}-t]|{\bf r},n,\mu\rangle\langle {\bf r'},n',\mu|
\end{equation}
where $\Sigma_{n}(\omega)$ is an energy-dependent self-energy and
called the coherent potential. Because of the absence of
translational symmetry of the system along the $z$-direction, this
self-energy depends on the layer number $n$. Furthermore, the
self-energy is spin independent, because the system is at a
paramagnetic temperature ($T\gg T_c$).

The physical properties of the real system can be obtained from
the configurationally averaged Green's function, $\langle
G\rangle_{\mathrm{av}}=\langle(\omega-H)^{-1}\rangle_{\mathrm{av}}$
which is replaced in the CPA with an effective medium Green's
function $\bar{G}=(\omega-H_{\mathrm{eff}})^{-1}$. A direct
consequence of this replacement is the fact that the effective
scattering of a carrier at the impurity site is zero, on average
\cite{Gonis}. The matrix elements of the effective Green's
function are calculated by the following Dyson equation
\cite{Saffar1}:
\begin{eqnarray}\label{5}
\bar{G}_{n,n'}(\mathbf{r},\mathbf{r}')&=&G^0_{n,n'}(\mathbf{r},\mathbf{r}')
+\sum_{n''=1}^{N_z}\sum_{\mathbf{r}''}G^0_{n,n''}(\mathbf{r},\mathbf{r}'')\nonumber\\
&&\times\Sigma_{n''} \bar{G}_{n'',n'}(\mathbf{r}'',\mathbf{r}')\ ,
\end{eqnarray}
where the variable $\omega$ has been suppressed for simplicity.
Here, $G^0(\omega)$ is the clean system Green's function and its
matrix element is given by \cite{Saffar1}:
\begin{equation}\label{6}
G^0_{n,n'}(\mathbf{r},\mathbf{r}')=\frac{a^2}{{4\pi^2}}\sum_{\ell=1}^{N_z}
\int_{\mathrm{1BZ}}d\mathbf{k}_\parallel\,
G^0_{n,n'}(\ell,\mathbf{k}_\parallel)
\,e^{i\mathbf{k}_\parallel\cdot(\bf{r}-\bf{r}')} \  .
\end{equation}
In this equation, $\ell$ denotes the subband number (mode), ${\bf
k}_{||}$ is a wave vector parallel to the layer, and the integral
is taken over all the wave vectors in the first Brillouin zone
(1BZ) of the two-dimensional lattice \cite{Saffar2}.
$G^0_{n,n'}(\ell,\mathbf{k}_\parallel)$ is the mixed Bloch-Wannier
representation of $G^0$ which can be expressed as
\begin{equation}\label{GLK}
G^0_{n,n'}(\ell,\mathbf{k}_\parallel;\omega)=\frac{h_{n,n'}(\ell)}{\omega+i\eta-\varepsilon_{\ell}(\mathbf{k}_\parallel)}\
,
\end{equation}
where
$h_{n,n'}(\ell)=\frac{2}{(N_z+1)}\sin(\frac{\ell\pi}{N_z+1}n)
\sin(\frac{\ell\pi}{N_z+1}n')$, $\eta$ is a positive infinitesimal
and $\varepsilon_{\ell}(\mathbf{k}_\parallel)$ is the electronic
band structure for mode $\ell$ \cite{Saffar2}.

To obtain the CPA condition, we define a perturbation potential
energy as
\begin{eqnarray}\label{v}
V&=&H-H_{\mathrm{eff}} \nonumber \\
&=&\sum_{{\bf r},n}v_{{\bf r},n} \ ,
\end{eqnarray}
where $v_{{\bf r},n}=v^A_{{\bf r},n}$  for the $A$ site and
$v_{{\bf r},n}=v^M_{{\bf r},n}$  for the $M$ site are given by the
following equations \cite{Saffar1}:
\begin{equation}\label{7}
v_{{\bf r},n}^A=\sum_{\mu}\mid {{\bf
r},n},\mu\rangle[\varepsilon_A-\Sigma_n]\langle {{\bf
r},n},\mu\mid\ ,
\end{equation}
\begin{equation}\label{8}
v_{{\bf r},n}^M=\sum_{\mu\nu}\mid {{\bf
r},n},\mu\rangle[(\varepsilon_M-\Sigma_n)\delta_{\mu,\nu}-I{\mathbf
S}_{{\bf r},n}{\bf\cdot{\mathbf \sigma}}_{\mu\nu}]\langle {{\bf
r},n},\nu\mid\ .
\end{equation}
Substitution of a fraction $x$ of the element $A$ by magnetic
impurity atom $M$ will include both substitutional disorder and
spin scattering, hence we can expect a semiconductor with magnetic
properties. In such a case, the CPA equation for the system is
expressed as:
\begin{equation}\label{cpa}
\langle t_{{\bf r},n}\rangle_{av}=(1-x)t^A_{{\bf r},n}+x\langle
t^M_{{\bf r},n}\rangle_{spin}=0\  ,
\end{equation}
where $t^A_{{\bf r},n}$ ($t^M_{{\bf r},n}$) represents the
complete scattering associated with the isolated potential
$v^A_{{\bf r},n}$ ($v^M_{{\bf r},n}$) due to the $A(M)$ atom in
the effective medium and $\langle\cdots\rangle_{spin}$ denotes
average over the spin scattering at the $M$ site \cite{Saffar1}.
At the paramagnetic temperature the orientation of localized spin
is completely random; hence the probability of each state is
$1/2$. Therefore, the associated $t$-matrices for an arbitrary
site ${\bf r}$ in the $n$th atomic layer are given by:
\begin{equation}\label{ta}
t^A_{{\bf r},n}=\frac{\varepsilon_A-\Sigma_n(\omega)}
{1-(\varepsilon_A-\Sigma_n(\omega))F_n(\omega)}\  ,
\end{equation}
and
\begin{eqnarray}\label{tm}
t^M_{{\bf
r},n}&=&\frac{1}{2}\left[\frac{\varepsilon_M-IS-\Sigma_n(\omega)}
{1-(\varepsilon_M-IS-\Sigma_n(\omega))F_n(\omega)}\right]\\
\nonumber
&+&\frac{1}{2}\left[\frac{\varepsilon_M+IS-\Sigma_n(\omega)}
{1-(\varepsilon_M+IS-\Sigma_n(\omega))F_n(\omega)}\right]\  ,
\end{eqnarray}
where $F_n(\omega)=\bar{G}_{n,n}(\mathbf{r},\mathbf{r};\omega)$ is
the diagonal element of the effective Green's function matrix
\cite{Gonis,Saffar1}. From the CPA condition, Eq. (\ref{cpa}), one
can derive an equation for the self-energy of $n$th layer. Then,
using such an equation and also Eq. (\ref{5}) which gives a system
of linear equations, one can obtain self-consistently the
self-energy $\Sigma_n(\omega)$ and the Green's function
$F_n(\omega)$ in each layer. Then, the LDOS per site in the $n$th
layer, $g_n(\omega)$, is calculated by
\begin{equation}\label{ldos}
g_n(\omega)=-\frac{1}{\pi}\,\mathrm{Im}\,F_n(\omega).
\end{equation}
On the other hand, to calculate the optical absorption spectrum,
we assume that both the magnetic and nonmagnetic atoms have equal
transition dipole moments. Therefore, when the $\ell$th subband is
optically excited, the layer-dependent optical absorption is given
by the contribution of ${\bf k_{\|}}=0$ component of the
electronic states that is related to the $\Gamma$ point in the 1BZ
\cite{Saffar1}. Accordingly, the optical absorption of the $n$th
layer, due to the creation of an exciton in the system, can be
defined as
\begin{equation}\label{optic}
A_n(\omega)=-\frac{1}{\pi}\,\mathrm{Im}\sum_{\ell=1}^{N_z}\bar{G}_{n,n}(\ell,\mathbf{k}_\parallel=0;\omega)\
.
\end{equation}
Note that, the calculation of $A_n(\omega)$ is not restricted to
the lowest $\ell=1$ subband, because the numerical results showed
that the contribution of $\mathbf{k}_\parallel=0$ component in the
other subbands, i.e. $\ell$=2,3,4 and 5, is finite and hence, all
the subband contributions must be included. In the next section we
present the numerical results of LDOS and optical absorption
spectrum for the system under consideration.
\begin{figure}
\centerline{\includegraphics[width=0.8\linewidth]{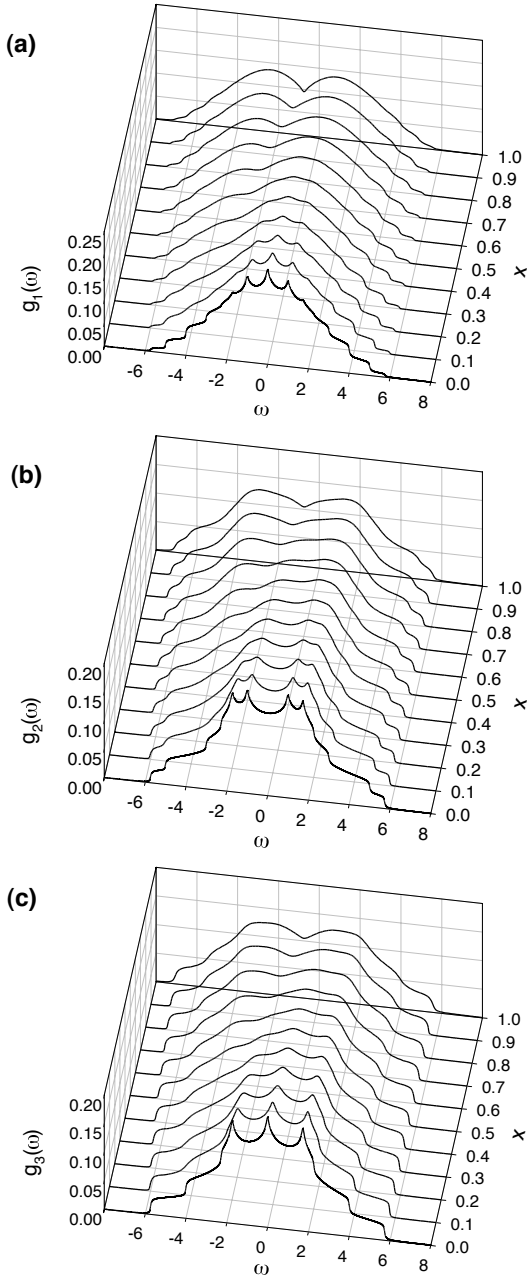}}
\caption{The LDOS for a DMS multilayer with $\varepsilon_M=
-0.75\,t$ and $IS=-1.5\,t$ as functions of energy $\omega$ and
alloy concentration $x$ for the atomic layers (a) $n=1,5$, (b)
$n=2,4$, and (c) $n=3$, respectively. $\omega$ is measured in
units of t.}
\end{figure}

\begin{figure}
\centerline{\includegraphics[width=0.8\linewidth]{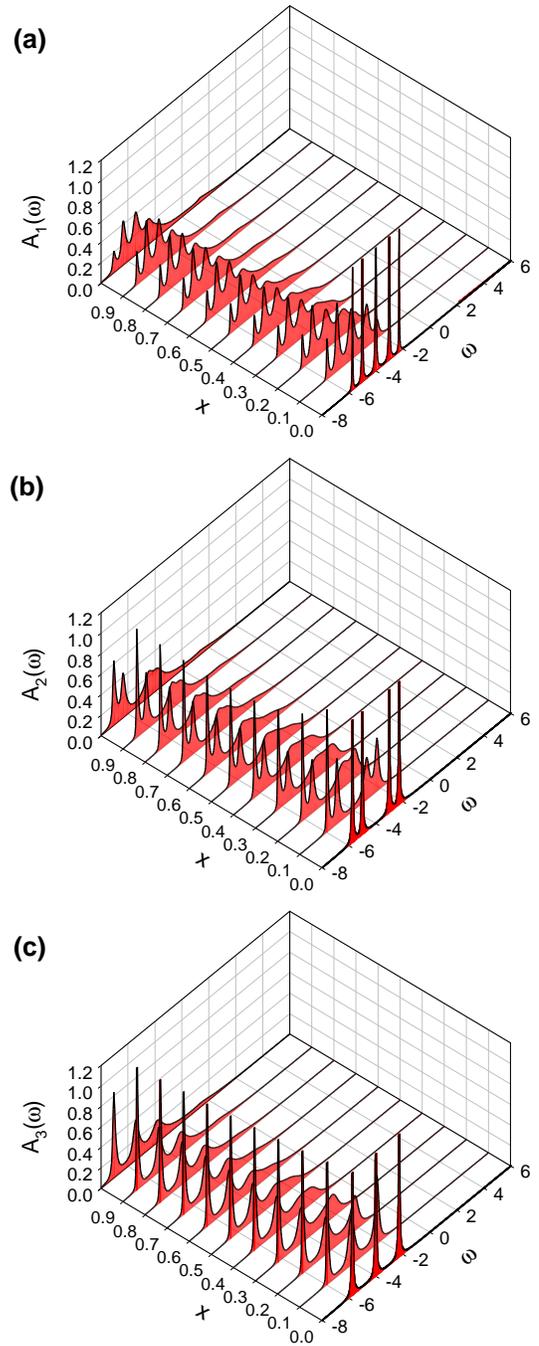}}
\caption{The optical absorption spectrum for a DMS multilayer with
$\varepsilon_M= -0.75\,t$ and $IS=-1.5\,t$ as functions of energy
$\omega$ and alloy concentration $x$ for the atomic layers (a)
$n=1,5$, (b) $n=2,4$, and (c) $n=3$, respectively. $\omega$ is
measured in units of t.}
\end{figure}
\section{Results and discussion}
Based on the above formalism we study the electronic states and
the optical absorption spectrum in a DMS alloy multilayer
consisting of five atomic layers, i.e. $N_z=5$. All the energies
are measured in units of $t$ and we set $\varepsilon_A$=0 as an
origin of the energy. We perform the calculations for two sets of
weak ($\varepsilon_M=-0.75\,t$, $IS=-1.5\,t$) and strong
($\varepsilon_M=-4.0\,t$, $IS=-3.5\,t$) interactions. It has been
shown that, the band gap opening in DMSs depends on the value of
magnetic ion chemical (or spin independent) potential and the
strength of exchange interaction \cite{Takahashi2}. Therefore, by
choosing the above values for $\varepsilon_M$ and $IS$ we can
model the system to show gap opening in the electronic states due
to the doping of $M$ atoms into the $AB$ structure. Since the
system consists of five atomic layers, one can expect five energy
subbands in the band structure of the system \cite{Saffar1}. Each
subband is attributed to one of the atomic layers. In addition,
due to the geometrical symmetry of the system in the $z$-
direction, the electronic states and hence the optical properties
in the layers $n=1$ and 5 and also in the layers $n=2$ and 4 are
the same. Therefore, we only present the LDOS and the optical
absorption spectrum for the layers $n=1$, 2 and 3.

\begin{figure}
\centerline{\includegraphics[width=0.8\linewidth]{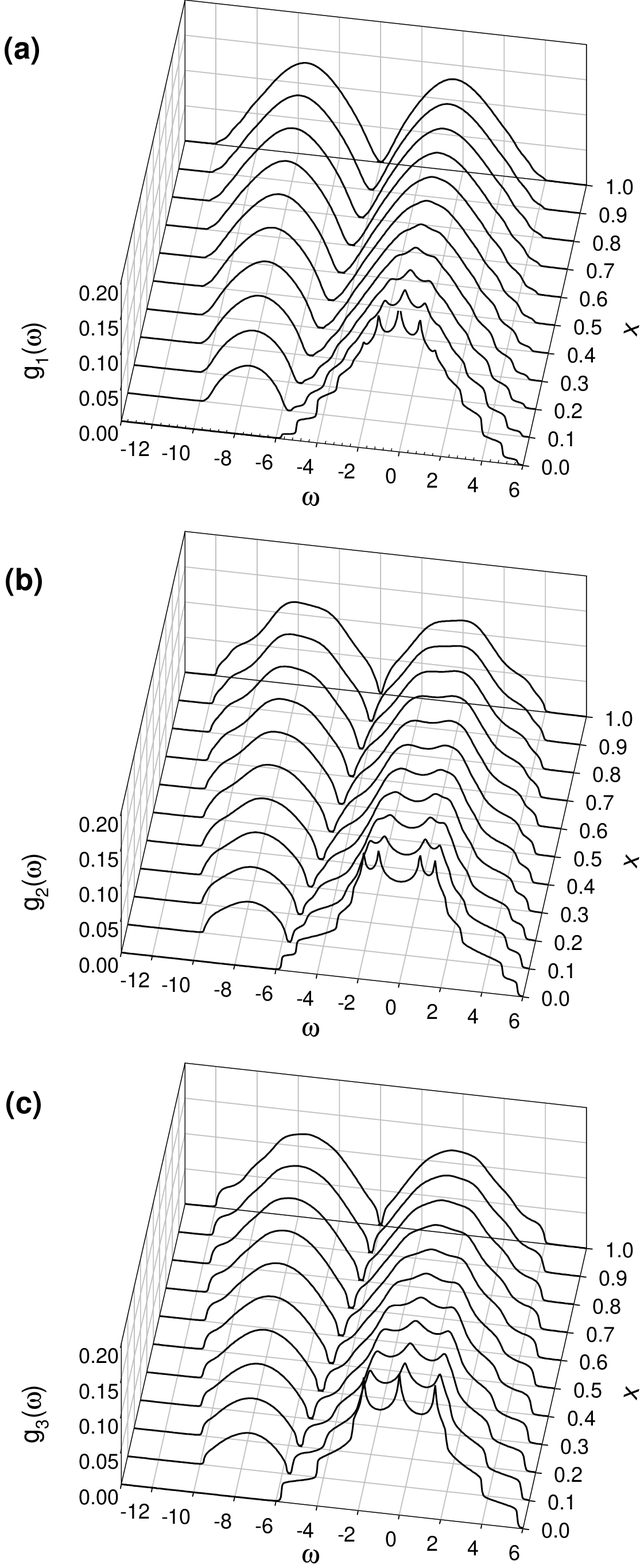}}
\caption{The same as Fig. 1 but for $\varepsilon_M= -4.0\,t$ and
$IS=-3.5\,t$.}

\end{figure}
\begin{figure}
\centerline{\includegraphics[width=0.8\linewidth]{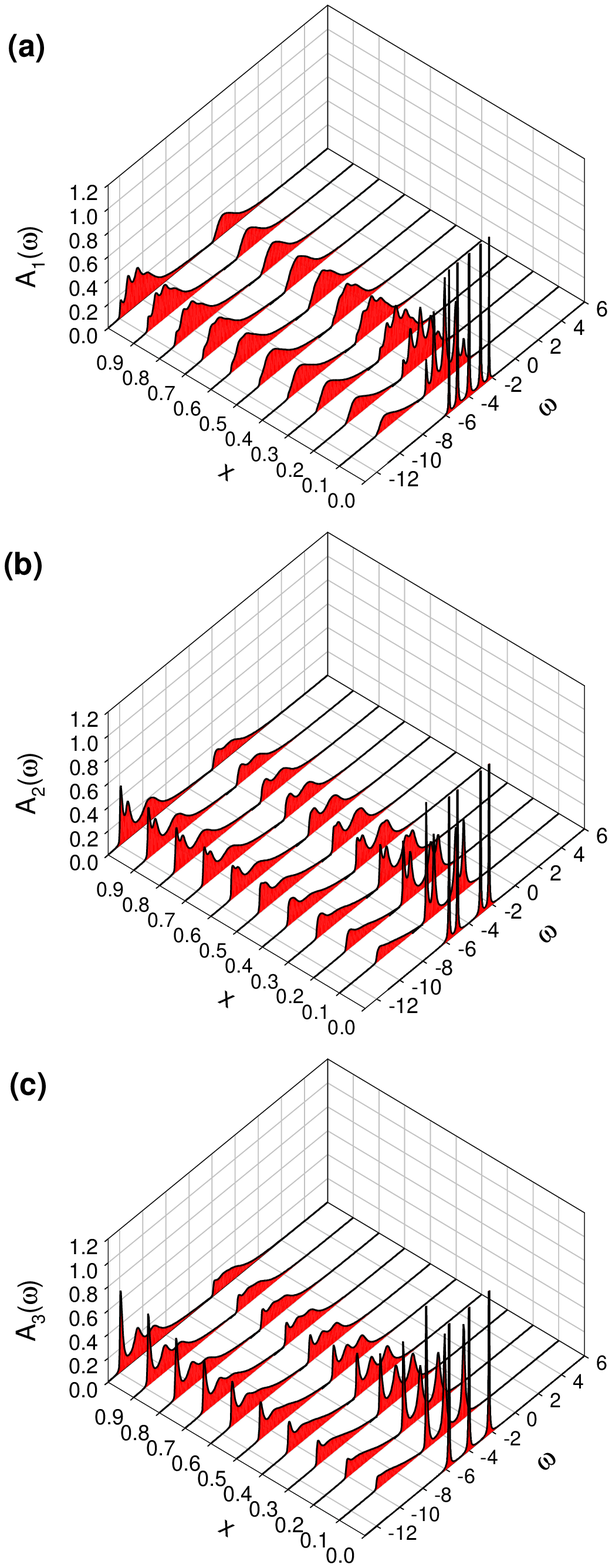}}
\caption{The same as Fig. 2 but for $\varepsilon_M= -4.0\,t$ and
$IS=-3.5\,t$.}
\end{figure}
Fig. 1 shows the dependence of electronic states on the carrier
energy $\omega$ and the alloy concentration $x$ in the case of
weak interaction, i.e., $\varepsilon_M=-0.75\,t$, $IS=-1.5\,t$.
Because of the 2D nature of the atomic layers of the system, one
can see the van Hove singularities and steplike features in the
LDOS which strongly depend on the magnetic ion concentration. The
energy position of these features is different in various layers
due to the quantum size effects which cause different electronic
states in different layers. In the case of $x=0$ the system is in
completely nonmagnetic case, i.e., the original $AB$ system, while
in the case of $x=1$, the $AB$ system changes to the $MB$ system
that is a clean magnetic system. When the alloy concentration is
increased, the bottom of the bands shifts towards lower energies
while the top of the bands remains approximately at a fixed
energy. The shift of the bands is due to the enhancement in the
spin fluctuation of $M$ ions with increasing $x$. In the limit of
$x\longrightarrow 1$, a dip appears close to the center of the
band that for $x=1$, its energy position is at
$\varepsilon=-0.75\,t$, i.e. at the on-site energy of the $M$
ions. The optical absorption spectrum, associated with the above
electronic states, is shown in Fig. 2. The results show sharp
features in the optical spectra corresponding to some of the
features seen in the LDOS at low energies. These features are
remarkable in the case of the clean system, i.e. in the $AB$
system. When the magnetic ions are introduced, that is in the case
of alloy system, the absorption peaks are broadened and the height
of sharp features decreases. Because of the dependence of
$A_n(\omega)$ to the electronic states through the imaginary part
of Green's function, one can see a shift in the optical spectra
toward lower energies, similar to that of the electronic states.
An important feature in Fig. 2 is the dependence of optical
absorption spectrum on the layer number. This means that the
photon absorption in low dimensional systems is different in
various layers. In addition, one can find that in the case of weak
interaction no energy gap appears in the optical spectrum.
Accordingly, the density of states and the absorption spectrum in
the weak interaction indicate that the $A$ and $M$ atomic states
are amalgamated into a single band when the exchange coupling is
small.

To study the effect of strong exchange interaction on the
electronic and optical properties of the system, we have shown in
Fig. 3 and 4 the layer-dependent of LDOS and optical absorption
for $\varepsilon=-4.0\,t$ and $IS=-3.5\,t$. The influence of
magnetic impurity causes an impurity band at the bottom of the
host band, whose bandwidth depends on the $M$ concentration $x$.
With increasing $x$, the contribution of spin scattering by the
localized spins in the scattering process of the carriers
increases and the impurity band and hence, the whole electronic
spectrum is broadened. This broadening which does not appear in
LDOS of the NMS alloys \cite{Saffar2} is a consequence of the
localized spin fluctuation of magnetic impurities. It is evident
that the strong exchange interaction splits the host band into two
subbands. We see that, the behavior of the electronic states of
each impurity band completely depends on the magnetic ion
concentration. For $x$=1, the two bands are symmetric and
correspond to the parallel coupling and antiparallel coupling
between the carrier spin and the localized spin, in agreement with
the results of the CPA for ordinary magnetic semiconductors at a
paramagnetic temperature \cite{Takahashi2,Rangette}. The energy
position of the electronic band center is located at the energy
$\varepsilon=-4.0\,t$, i.e. at the on-site energy of the $M$ ions,
which is a consequence of the classical spin treatment of the
localized spins at a paramagnetic temperature. Note that, the
quantum effect of the localized spins introduces some further
complications in the results and the band symmetry is broken.

One of the main differences between the electronic states in the
magnetic and nonmagnetic semiconductor alloys is the variation of
$g_n(\omega)$ with increasing $x$. In the CPA method and in the
absence of magnetic impurities \cite{Saffar2}, a NMS alloy for
$x=1$ is converted to the same original NMS but with an energy
shift in the electronic states. This means that, the behavior of
LDOS with energy variation at $x=1$ is the same behavior of the
LDOS at $x=0$. By doping a NMS with magnetic impurities, however,
the behavior of LDOS might be completely different with the
electronic states of the original undoped system, specially at
$x=1$. This difference appears because of the multiple scattering
effect due to the exchange interaction between the carrier spin
and the localized spin of magnetic ion, which is most remarkable
in both cases of $x=0$ and $x=1$. Because of the dependence of
optical spectrum on the electronic states, this remarkable
difference for the cases of $x=0$ and $x=1$ can be clearly seen in
this study. It is important to note that, if we set $IS$=0.0, the
present formalism will give the same results for a NMS multilayer
in which there is no band broadening in the electronic spectra
\cite{Saffar2}.
\begin{figure}
\centerline{\includegraphics[width=0.8\linewidth]{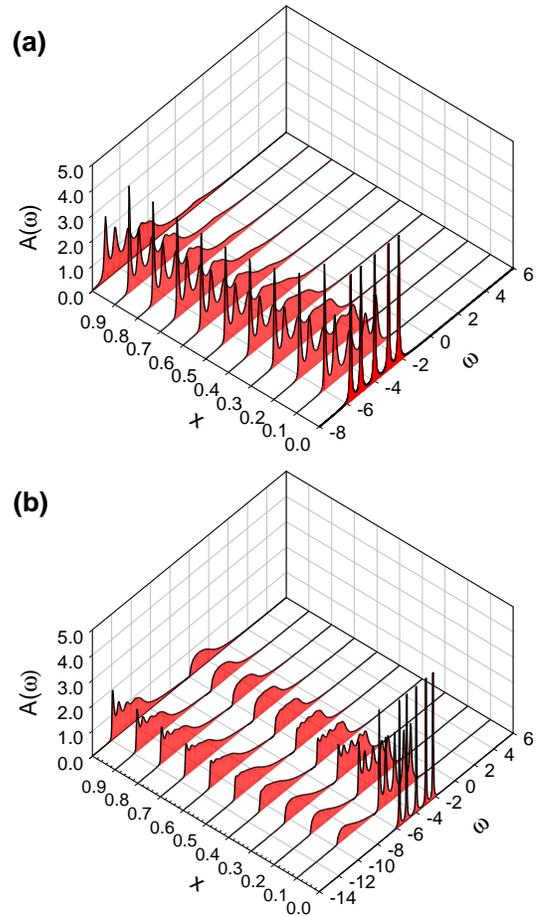}}
\caption{The total optical absorption spectrum for the two DMS
multilayers as a function of energy $\omega$ and alloy
concentration $x$. In (a) $\varepsilon_M= -0.75\,t$ and
$IS=-1.5\,t$ and in (b) $\varepsilon_M= -4.0\,t$ and $IS=-3.5\,t$.
$\omega$ is measured in units of $t$.}
\end{figure}

The electronic band splitting in the strong exchange interaction
regime creates the two subbands in the optical absorption
spectrum. For $0<x<1$, the lower energy band corresponds to the
case that the carrier spin is parallel with the localized spin,
while the higher band belongs to the $A$ atom \textit{and} the
case in which the carrier spin is antiparallel with the localized
spin. Therefore, such an optical absorption spectrum might
correspond to the persistence type \cite{Takahashi}. For $x=1$,
the spectrum completely belongs to the $M$ atoms, i.e. the clean
magnetic system. Optical band broadening in the all layers can be
clearly seen in the spectrum due to the chemical and magnetic
disorder. Furthermore, the bands show a sharp feature at the
optical band edge, depending on the value of $x$. The strength of
this feature increases for the interior layers and indicates that
most of the photon energy can be absorbed by the interior layers
of the system. Roughly, for $x\leq 0.5$, the sharp features
appears at the bottom of the higher band, while for $x>0.5$ the
respective features appear at the bottom of the lower band.  Note
that, the band broadening, which is accompanied by energy shift of
the bottom of the optical bands, demonstrates the operation of a
magnetic semiconductor multilayer as an optical device which can
be tuned at desired wavelength.

Another set of the results that we present here is the bulk
(total) optical absorption for the weak and strong exchange
interactions. Due to the fact that the optical absorption of each
layer is not experimentally measurable yet, we have calculated the
total magnitude of this quantity here as well (see Fig. 5). For
this purpose, we make a summation over all five layers to obtain
the total optical absorption of the system, i.e.
$A(\omega)=\sum_{n=1}^{N_z}A_n(\omega)$. All features that we see
in Figs. 5(a) and (b) are compatible with those of the individual
layers. It is clear that the sharp features of the amalgamation
type (Fig. 5(a)) at the bottom of the optical band are much
stronger than those of the persistence type (Fig. 5(b)). The
results also show that the optical absorption of a DMS multilayer
has an obvious broadening in comparison with the NMS multilayers
\cite{Saffar2}, and in the case of larger $IS$ the amount of
broadening is higher and a gap opening occurs. Therefore, the
transition from the amalgamation type to the persistence type in
the DMS multilayers depends on the strength of the exchange
coupling $IS$, and the magnetic ion concentration $x$.

\section{Conclusion}
We have studied the effects of magnetic impurity on the electronic
and optical properties of a semiconductor multilayer under the
assumption that the transition dipole moments of the magnetic and
nonmagnetic ions are the same. Using the single-site CPA for a
random distribution of the impurity atoms, we investigated the
influence of exchange interaction strength and the impurity
concentration on the LDOS and the optical absorption spectra of
the system. In such multilayers, the magnetic impurities shift the
spectra towards lower energy-side, broaden the bands, and strongly
affect on the sharp features of the spectra, even in the case of
weak exchange coupling. In the case of strong exchange coupling,
an energy gap opens in the electronic and optical bands. The
interior layers show a higher contribution in the optical
absorption process of the system relative to the surface layer,
together with the fact that the amalgamation type alloys in
comparison with the persistence type alloys need a higher amount
of photons energy to make an excitation. These results might be
helpful for the development of DMS multilayers in magneto-optical
devices at a paramagnetic temperature.

\end{document}